\documentclass[preprint,12pt]{elsarticle}

\usepackage[ruled,vlined,linesnumbered]{algorithm2e}
\usepackage{times}
\usepackage{setspace}
\usepackage{amsmath}
\usepackage{amsfonts}
\usepackage{placeins}
\usepackage{caption}
\usepackage{amsthm}
\usepackage{booktabs}
\usepackage{array}
\usepackage{textcomp}
\usepackage{longtable}
\usepackage{stfloats}
\usepackage{url}
\usepackage{verbatim}
\usepackage{graphicx}
\usepackage{cite}
\usepackage{xcolor}
\usepackage{setspace}
\usepackage{epstopdf}
\usepackage{bm}
\usepackage{float}
\usepackage{amssymb}
\usepackage{microtype}
\usepackage{algorithmic}
\usepackage{multirow}
\usepackage{pifont}
\newcommand{\cmark}{\ding{51}}
\newcommand{\xmark}{\ding{55}}
\allowdisplaybreaks

\journal{Applied Energy}

\begin{document}
\begin{frontmatter}
\title{Real-Time Coordinated Operation of Off-Grid Wind Powered Multi-Electrolyzer Systems Considering Thermal Dynamics and HTO Safety}

\author[1,2,3]{\textcolor{black}{Chang Su}}
\author[1,2,3]{\textcolor{black}{Ming Li}}
\author[1,2,3]{\textcolor{black}{Zhanglin Shangguan}}
\author[1,2,3]{\textcolor{black}{Zhaojian Wang}}
\author[1,2,3]{\textcolor{black}{Bo Yang}\corref{cor1}}
\cortext[cor1]{Corresponding author}
\ead{bo.yang@sjtu.edu.cn}

\affiliation[1]{organization={State Key Laboratory of Submarine Geoscience, School of Automation and Intelligent Sensing, Shanghai Jiao Tong University},
city={Shanghai},
postcode={200240},
country={China}}

\affiliation[2]{organization={Key Laboratory for System Control and Information Processing, Ministry of Education of China},
city={Shanghai},
postcode={200240},
country={China}}

\affiliation[3]{organization={Shanghai Engineering Research Center of Intelligent Control and Management},
city={Shanghai},
postcode={200240},
country={China}}

\begin{abstract}
Coordinated operation of alkaline water electrolysis (AWE) systems with multiple electrolyzers under fluctuating renewable power input is challenging due to varying power availability and dynamic safety constraints. Moreover, the conventional separation between optimization and control may result in inconsistent decisions across timescales. To address these issues, this paper proposes a two-layer coordinated operation method integrating feedback optimization (FO) with a projection-based safety layer. The FO layer generates real-time reference inputs to improve renewable energy utilization, while the safety layer corrects these inputs to ensure compliance with operational and safety constraints. To explicitly address the safety constraints arising from the inertial dynamics of AWE systems, discrete-time control barrier function theory is incorporated into the safety layer, thereby enhancing safety assurance and online computational tractability. Theoretical analysis establishes the feasibility and effectiveness of the proposed method. Case studies based on annual wind generation data show that the proposed method achieves high energy utilization, maintains safe operation, and demonstrates online applicability, scalability, and robustness.
\end{abstract}

\end{frontmatter}

\section*{Nomenclature}
\noindent\textbf{Parameters and variables}
\begin{list}{}{%
\setlength{\labelwidth}{1.2cm}%
\setlength{\labelsep}{0.15cm}%
\setlength{\leftmargin}{1.35cm}%
\setlength{\itemsep}{0pt}%
\setlength{\parsep}{0pt}%
\setlength{\topsep}{2pt}%
\renewcommand{\makelabel}[1]{\makebox[\labelwidth][l]{#1}}}
\item[$u_{ele}$] Operating voltage of electrolyzer (V)
\item[$i_{ele}$] Operating current of electrolyzer (A)
\item[$P_{ele}$] Electrolyzer power consumption (W)
\item[$u_{rev}$] Reversible voltage (V)
\item[$T_{ele}$] Electrolyzer temperature ($^{\circ}\mathrm{C}$)
\item[$N_{cell}$] Number of cells
\item[$u_{limit}$] Voltage limit of a single cell (V)
\item[$C_{th}$] Equivalent thermal capacitance (J/K)
\item[$R_{th}$] Equivalent thermal resistance (K/W)
\item[$T_a$] Ambient temperature ($^{\circ}\mathrm{C}$)
\item[$\dot n_{cross}$] Total hydrogen crossover rate to the anode (mol/s)
\item[$\dot n_{diff}$] Diffusion induced hydrogen crossover rate (mol/s)
\item[$\dot n_{con}$] Convection induced hydrogen crossover rate (mol/s)
\item[$\dot n_{lye}$] Electrolyte circulation induced hydrogen crossover rate (mol/s)
\item[$n$] Molar amount (mol)
\item[$\dot n$] Molar flow rate (mol/s)
\item[$v_{lye}$] Electrolyte flow rate (m$^3$/s)
\item[$V_{an}$] Anode compartment volume (m$^3$)
\item[$V_{sep,g}$] Separator gas-phase volume (m$^3$)
\item[$\tau_{sep,l}$] Separator time constant (s)
\item[$P_{wind}$] Available wind power (W)
\item[$N_{ele}$] Number of electrolyzers
\item[$T_{wind}$] Length of the wind power horizon
\item[$\Delta t$] Sampling period (s)
\item[$\Delta i_{max}$] Maximum current ramp rate (A/s)
\end{list}

\noindent\textbf{Superscripts and subscripts}
\begin{list}{}{%
\setlength{\labelwidth}{1.2cm}%
\setlength{\labelsep}{0.15cm}%
\setlength{\leftmargin}{1.35cm}%
\setlength{\itemsep}{0pt}%
\setlength{\parsep}{0pt}%
\setlength{\topsep}{2pt}%
\renewcommand{\makelabel}[1]{\makebox[\labelwidth][l]{#1}}}
\item[$ele$] Electrolyzer
\item[$an$] Anode compartment
\item[$sep$] Separator
\item[$l$] Liquid phase
\item[$g$] Gas phase
\item[$in$] Inflow
\item[$out$] Outflow
\end{list}

\section{Introduction}\label{S1}
Renewable hydrogen production powered by wind energy is increasingly regarded as an important pathway for improving renewable energy utilization and supporting the transition to low carbon energy systems \citep{P1}. Among the available electrolysis technologies, alkaline water electrolysis (AWE) is widely considered a promising option for large scale hydrogen production because of its high technological maturity, relatively low cost, and suitability for industrial scale deployment \citep{P2,P3}. However, when directly supplied by off-grid wind power, a single AWE unit struggles to accommodate wind power fluctuations while maintaining efficient and safe operation, owing to its restricted low-load capability and dynamic constraints associated with thermal behavior and hydrogen in oxygen concentration (HTO) \citep{P4}. In practical applications, wind powered hydrogen production is therefore typically realized by multiple parallel electrolyzers to enhance operational flexibility \citep{P5,P6}. Nevertheless, coordinated operation remains challenging because each unit is constrained by transient thermal evolution, ramping capability, HTO safety requirements, and instantaneous operating limits \citep{P7,P8,P9,P10}, while multiple units are further coupled through system level objectives such as renewable power utilization and hydrogen production performance \citep{P11,P12}.

Existing studies on this topic can be broadly classified into two main categories, namely rule-based methods and optimization-based methods. Rule-based methods have been widely investigated for the coordinated operation of multi-electrolyzer systems, owing to their simple structure, straightforward implementation, and relatively low computational complexity. In \citep{P13}, two strategies, namely the equal force strategy and the doctor triage strategy, are proposed for multi-electrolyzer systems. The equal force strategy distributes power among electrolyzers according to a fixed sequence, while the doctor triage strategy incorporates a rotation mechanism to alleviate the operating imbalance among different electrolyzers. In \citep{P14}, the minimum operating power threshold of electrolyzers is further considered, and two corresponding strategies, namely the piecewise equal distribution strategy and the piecewise equal cycling distribution strategy, are proposed. In \citep{P15}, the switching constraints of electrolyzers are taken into account, and a start-stop strategy is developed for wind-hydrogen systems to regulate electrolyzer operation under fluctuating wind power. \citep{P16} proposes a two-phase coordination strategy, in which rapid start-up and subsequent equal power allocation are combined to improve renewable energy utilization and extend electrolyzer lifespan. Overall, rule-based methods enable real-time coordination with low implementation complexity. However, these methods rely heavily on manually designed rules and are limited in handling coupled dynamic and safety-critical constraints, especially when transient thermal behavior and HTO safety requirements must be explicitly considered.

In addition to rule-based methods, optimization-based methods have also been widely investigated for the coordinated operation of multi-electrolyzer systems. In this class of methods, dispatch and scheduling decisions are typically determined within an optimization framework that explicitly accounts for system level objectives and multiple operating constraints. In \citep{P17}, a source and load collaborative optimization strategy is proposed for multi-electrolyzer arrays to improve hydrogen production efficiency, enhance operational balance, and reduce cold starts. In \citep{P18}, a voltage degradation based optimization strategy is developed, where temperature, HTO impurities, and cumulative voltage degradation are incorporated into the power allocation framework to promote safe and balanced operation and prolong system lifetime. In \citep{P19}, an optimal production scheduling approach is proposed for utility-scale P2H plants with multiple AWEs, in which dynamic thermal and HTO impurity effects are integrated into the scheduling model to coordinate state transitions and load allocation. In \citep{P20}, a rolling optimization based operational strategy is proposed, where efficiency variation, start-stop processes, thermal management, and degradation are jointly considered to improve system efficiency and electrolyzer lifetime performance. Overall, optimization-based methods provide a systematic framework for coordinating multi-electrolyzer systems and handling coupled constraints and objectives. However, explicitly incorporating transient thermal behavior and dynamically evolving HTO safety constraints often results in high computational complexity. In addition, many existing methods rely on renewable power forecasts, and their complexity increases significantly with the number of electrolyzers and the optimization horizon.

Although the aforementioned methods have advanced the coordinated operation of multi-electrolyzer systems, several important challenges still remain: (1) Existing coordinated operation studies are commonly based on steady state formulations, which limits their ability to accurately capture the transient response and dynamic operating characteristics of electrolyzers, thereby reducing the accuracy of power allocation decisions and potentially compromising operational safety and renewable energy utilization; (2) Upper-level power allocation or scheduling is commonly designed without explicitly accounting for the dynamic feasibility of the lower-level control system, which may lead to inconsistency between optimization decisions and actual closed-loop operation; (3) Achieving the strict satisfaction of safety-critical HTO constraints while maintaining computational tractability and scalability remains a significant challenge, especially for large-scale systems with dynamically evolving admissible operating regions. Table \ref{tab1} summarizes the comparison between the existing methods and the proposed method, highlighting the main methodological differences and the research gaps addressed in this work.

To address the above limitations, this paper proposes an optimization-control coordinated operation framework for wind powered multi-electrolyzer hydrogen production systems. Specifically, feedback optimization is adopted to realize real-time coordinated power allocation, while a projection-based safety layer is designed to guarantee the satisfaction of operational constraints during online implementation. The main contributions of this paper are summarized as follows:
\begin{itemize}
\item{A two-layer optimization-control coordinated operation framework is established for off-grid wind powered multi-electrolyzer systems. In the upper layer, feedback optimization generates real-time coordinated power allocation references to improve renewable power utilization, whereas in the lower layer, these references are adjusted online under dynamic operational constraints to ensure implementability. This structure coordinates reference generation and constrained implementation within a consistent timescale, thereby improving the consistency between power allocation decisions and actual operation.}
\item{A discrete-time control barrier function (CBF) is introduced to rigorously handle the HTO safety constraint governed by inertial internal states. By transforming the safety-critical state constraint into an input admissibility condition, the proposed method guarantees strict safety satisfaction at all sampling instants and improves the computational tractability and efficiency of online safety enforcement.}  
\item{Theoretical analysis establishes the feasibility of the projection-based safety layer, forward invariance of the HTO safe set, and uniform ultimate boundedness of the power mismatch. Numerical studies based on real wind power data further verify effective wind power utilization, strict satisfaction of safety constraints, real-time applicability, and scalability, while sensitivity analysis confirms robustness and provides guidance for parameter tuning.}
\end{itemize}

The remainder of this paper is organized as follows: Section \ref{S2} presents the electrolyzer models and formulates the coordinated operation optimization problem. Section \ref{S3} develops the proposed two-layer coordinated operation framework. Section \ref{S4} provides the theoretical analysis of the proposed framework. Section \ref{S5} presents numerical studies to verify its effectiveness. Section \ref{S6} concludes the paper.

\begin{table*}[t]
\centering
\caption{Comparison of the proposed strategy with previous literature.}
\label{tab1}
\renewcommand{\arraystretch}{1.15}
\setlength{\tabcolsep}{4.5pt}
\resizebox{\textwidth}{!}{%
\begin{tabular}{lccccccccc}
\toprule
\textbf{Key point of comparison} 
& \textbf{[13]} & \textbf{[14]} & \textbf{[15]} & \textbf{[16]} & \textbf{[17]}
& \textbf{[18]} & \textbf{[19]} & \textbf{[20]} & \textbf{This work} \\
\midrule
Rule-based
& \cmark & \cmark & \cmark & \cmark & \xmark & \xmark & \xmark & \xmark & \xmark \\

Optimization-based
& \xmark & \xmark & \xmark & \xmark & \cmark & \cmark & \cmark & \cmark & \cmark \\

Consider renewable power utilization rate
& \cmark & \cmark & \xmark & \cmark & \cmark & \cmark & \cmark & \cmark & \cmark \\

Consider multiple electrolyzers
& \cmark & \cmark & \cmark & \cmark & \cmark & \cmark & \cmark & \cmark & \cmark \\

Consider thermal dynamics
& \xmark & \xmark & \xmark & \xmark & \xmark & \cmark & \cmark & \cmark & \cmark \\

Consider HTO safe requirements
& \xmark & \xmark & \cmark & \xmark & \xmark & \cmark & \cmark & \xmark & \cmark \\

Depend on renewable power forecasting
& \xmark & \xmark & \xmark & \xmark & \cmark & \xmark & \cmark & \xmark & \xmark \\

Optimization-control timescale coordination
& \xmark & \xmark & \xmark & \xmark & \xmark & \xmark & \xmark & \xmark & \cmark \\

Validation under different cluster sizes
& \cmark & \cmark & \xmark & \xmark & \xmark & \xmark & \cmark & \xmark & \cmark \\

Real-time applicability
& \cmark & \cmark & \cmark & \cmark & \xmark & \cmark & \xmark & \cmark & \cmark \\
\bottomrule
\end{tabular}%
}
\end{table*}
\section{Electrolyzer modeling and problem formulation}\label{S2}
\begin{figure}[t]
	\centering  
	\includegraphics[scale=0.4]{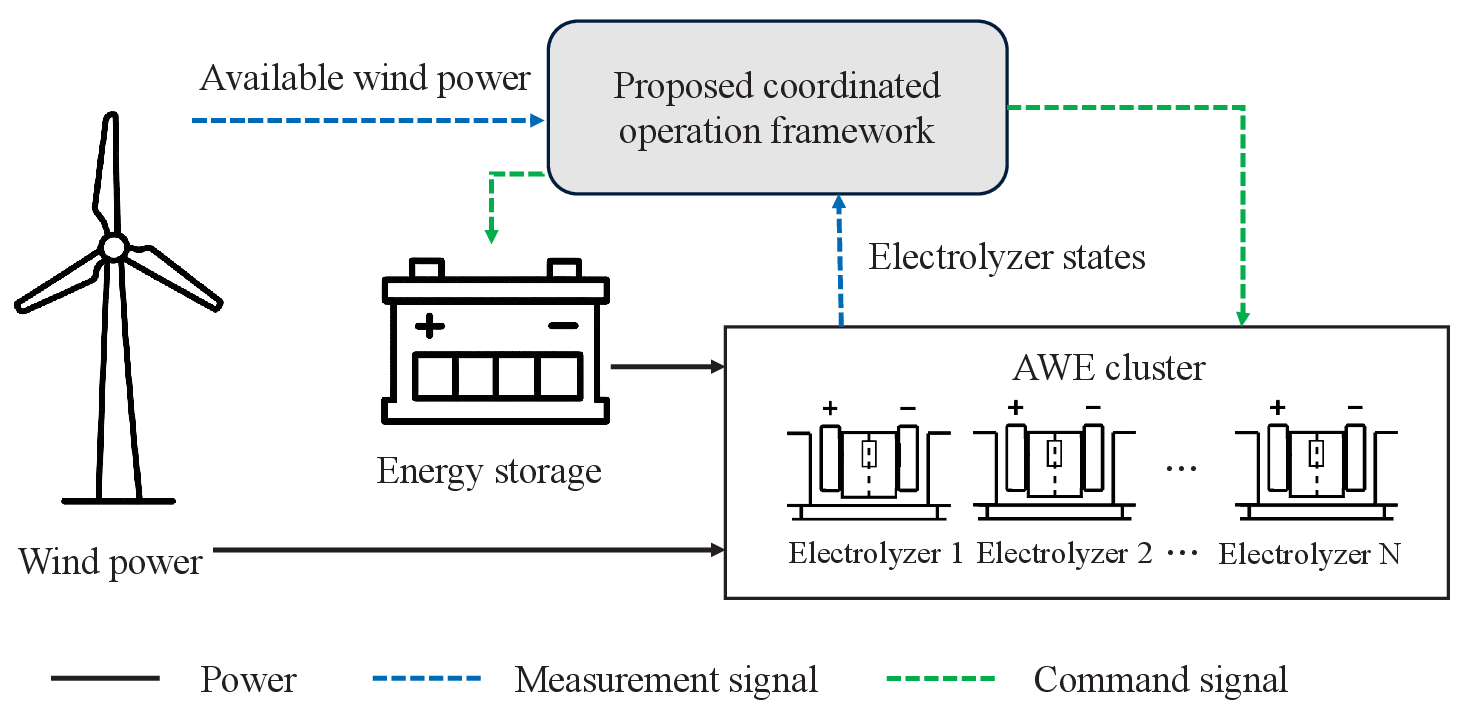}
	\caption{Schematic diagram of the proposed coordinated operation framework for the wind-storage-AWE system.} 
	\label{Figure 1} 
\end{figure}

This paper considers an off-grid wind powered hydrogen production system consisting of a wind power source, a battery energy storage unit, and a cluster of AWEs. The available electrical power is coordinated among the electrolyzers in real time to improve renewable energy utilization while ensuring safe operation. A schematic diagram of the considered system is shown in Fig. \ref{Figure 1}.

Due to the intermittency and variability of wind generation, the available power needs to be dynamically distributed among multiple electrolyzers. However, the operating limits of electrolyzers are constrained by their electrochemical characteristics, thermal dynamics, and safety requirements such as HTO concentration. Therefore, accurate models describing these characteristics are required to determine the feasible operating region of electrolyzers. The corresponding models are introduced below to support the subsequent power allocation problem.

\subsection{Power model}
Since the power consumption capability of an electrolyzer depends strongly on its operating temperature, the corresponding maximum allowable power determines its feasible operating region in the power allocation problem. Therefore, it is necessary to characterize the temperature-dependent maximum consumable power of the electrolyzer.
Among various first-principle models and empirical models, this paper adopts the semi-empirical model proposed in \citep{P21}, which explicitly describes the relationship between the maximum consumable power and temperature as follows:
\begin{equation}
{u_{ele}} = {u_{rev}} + \left( {{\rho _1} + {\rho _2}{T_{ele}}} \right){i_{ele}},
\label{E1}
\end{equation}
\begin{equation}
{u_{ele}} \le u_{ele}^{max} = {N_{cell}}{u_{limit}},
\label{E2}
\end{equation}
\begin{equation}
{i_{ele}} \le i_{ele}^{max} = ({{N_{cell}}{u_{limit}} - {u_{rev}}})/({{{\rho _1} + {\rho _2}{T_{ele}}}}),
\label{E3}
\end{equation}
where $\rho_1$ and $\rho_2$ are empirical parameters with $\rho_2 \le 0$. Accordingly, the maximum power that the electrolyzer can consume at a given temperature can be obtained as
\begin{equation}
P_{ele}^{max} = u_{ele}^{max}i_{ele}^{max}.
\label{E4}
\end{equation}
\subsection{Thermal model}
The temperature dynamics of the electrolyzer can be described using a lumped thermal model as follows \citep{P22}:
\begin{equation}
{C_{th}}\frac{{d{T_{ele}}}}{{dt}} = {P_{heat}} - {P_{diss}} - {P_{cool}}.
\label{E5}
\end{equation}

The heat generated by the electrolysis process, the heat dissipated to the ambient, and the heat removed by the cooling system are denoted by $P_{heat}$, $P_{diss}$, and $P_{cool}$, respectively, and are given by
\begin{equation}
{P_{heat}} = \left( {{\rho _1} + {\rho _2}{T_{ele}}} \right)i_{ele}^{2},
\label{E6}
\end{equation}
\begin{equation}
{P_{diss}} = \left( {{T_{ele}} - {T_a}} \right)/{R_{th}}.
\label{E7}
\end{equation}

Under the practical operating conditions considered in this paper, the electrolyzer temperature remains within the admissible range and does not reach the threshold for active cooling. Therefore, $P_{cool}$ is neglected in the thermal model.

\subsection{HTO model}\label{S2-c}
HTO concentration is a critical safety indicator for AWEs, because exceeding the prescribed safety threshold $2\%$ may pose an explosion risk \citep{P23}. In AWEs, hydrogen crossover mainly arises from diffusion, convection, and electrolyte circulation. Based on these mechanisms, the hydrogen crossover rate can be expressed as follows \citep{P24}:
\begin{equation}
{{\dot n}_{cross}} = {{\dot n}_{diff}} + {{\dot n}_{con}} + {{\dot n}_{lye}}.
\label{E8}
\end{equation}

According to \citep{P25}, hydrogen impurity is mainly accumulated in the gas phase of the separator, and the accumulation process can be described by a three-compartment model as
\begin{subequations}\label{E9}
\begin{equation}
{{\dot n}_{an}} = {{\dot n}_{an,in}} - {{\dot n}_{an,out}},
\end{equation}
\begin{equation}
{{\dot n}_{sep,l}} = {{\dot n}_{sep,l,in}} - {{\dot n}_{sep,l,out}},
\end{equation}
\begin{equation}
{{\dot n}_{sep,g}} = {{\dot n}_{sep,g,in}} - {{\dot n}_{sep,g,out}},
\end{equation}
where
\begin{equation}
{{\dot n}_{an,in}} = {{\dot n}_{cross}},
\end{equation}
\begin{equation}
{{\dot n}_{an,out}} = {{\dot n}_{sep,l,in}} = {n_{an}}{v_{lye}}/{V_{an}}/2,
\end{equation}
\begin{equation}
{{\dot n}_{sep,l,out}} = {{\dot n}_{sep,g,in}} = {n_{sep,l}}/{\tau _{sep,l}},
\end{equation}
\begin{equation}
\dot n_{sep,g,out}
=
\frac{n_{sep,g}\,\dot n_{O_2}}{P V_{sep,g}/(R T_{ele})},
\end{equation}
\end{subequations}
\begin{equation}
HTO = {{\dot n}_{sep,g,out}} / {{\dot n}_{{O_2}}},
\label{E10}
\end{equation}
where $P$, $R$, and $\dot n_{{O_2}}$ denote the pressure, the molar gas constant, and the oxygen molar production rate, respectively. 

\subsection{Problem formulation}
Based on the above models, an optimization problem is formulated to determine the power allocation strategy of off-grid electrolyzers powered by wind turbines. Considering the constraints of thermal effects, HTO safety requirements, and computational efficiency, the objective is to maximize energy utilization efficiency while ensuring safe system operation. The resulting optimization problem is defined as (P1). The goal is to develop a computationally efficient real-time solution framework for it. 
\begin{subequations}\label{E11}
\begin{align}
{\textrm{(P1)}}~~~~&\mathop {\min }\limits_{{i_{ele,i}}\left( t \right)} {\sum\limits_{t = 1}^{{T_{wind}}} {\left( {\sum\limits_{i = 1}^{{N_{ele}}} {{u_{ele,i}}\left( t \right){i_{ele,i}}\left( t \right)}  - {P_{wind}}\left( t \right)} \right)} ^2}
\label{E11a}\\
{\textrm{s.t.}}&0 \le \sum\limits_{i = 1}^{{N_{ele}}} {{P_{ele,i}}\left( t \right)}  \le {P_{wind}}\left( t \right),
\label{E11b}\\
&{u_{ele,i}}\left( t \right) \le u_{ele}^{max },
\label{E11c}\\
&{i_{ele,i}}\left( t \right) \le {i_{ele,max ,i}}\left( {{T_{ele,i}}\left( t \right)} \right),
\label{E11d}\\
&{P_{ele,i}}\left( t \right) \le {P_{ele,max ,i}}\left( {{T_{ele,i}}\left( t \right)} \right),
\label{E11e}\\
&\left| {{i_{ele,i}}\left( {t + 1} \right) - {i_{ele,i}}\left( t \right)} \right| \le \Delta {i_{max }},
\label{E11f}\\
&HT{O_i}\left( t \right) \le HT{O_{max }},
\label{E11g}
\end{align}
\end{subequations}
where \eqref{E11a} minimizes the power mismatch between wind generation and electrolyzer consumption. Constraint \eqref{E11b} defines the total power limit.
Constraints \eqref{E11c}, \eqref{E11d}, and \eqref{E11e} impose the voltage, current, and power limits of electrolyzers under given operating conditions, respectively.
Constraint \eqref{E11f} restricts the ramp rate of the electrolyzer current.
Constraint \eqref{E11g} ensures that the HTO concentration remains within safety limits, thereby guaranteeing safe hydrogen production.
Overall, problem (P1) aims to determine the power allocation of multiple electrolyzers such that wind power can be absorbed as much as possible while safe operation is ensured. 

Although problem (P1) is compactly formulated, it is essentially a dynamic constrained optimization problem, because the electrical, thermal, and HTO related variables evolve according to nonlinear dynamics and couple the input decisions over time. In particular, the HTO constraint is safety-critical, since exceeding the prescribed limit may lead to serious safety risks. Under fluctuating wind power, coordinated operation requires rapid updates to track the available renewable power while respecting dynamic safety limits. However, solving problem (P1) online at a short interval leads to a high computational burden, whereas a slower update may allow the system states to evolve significantly between successive optimization steps, so that the computed allocation may no longer remain implementable and may even become dynamically infeasible. Therefore, directly solving problem (P1) online makes it difficult to simultaneously achieve fast responsiveness, computational tractability, and operational safety. To address these challenges, a control-oriented real-time solution framework is developed in the next section.
\section{Control-oriented real-time solution framework}\label{S3}
To enable computationally efficient real-time implementation, a control-oriented solution framework is developed for problem (P1). Specifically, the power tracking objective is approximately handled through a feedback optimization law \citep{P26}, while the operational and safety constraints are enforced by a projection-based safety layer.
For sampled data implementation, the continuous-time plant dynamics are sampled with period $\Delta t$. 

\subsection{Feedback optimization-based power allocation}\label{S3-1}
In order to generate the desired power allocation input in real time, this subsection focuses on the power tracking objective in problem (P1), while the operational and safety constraints are addressed in the subsequent subsections through a projection-based safety layer. Since the electrolyzer temperature directly affects its power consumption capability, the temperature dynamics are retained in the feedback optimization layer. The HTO dynamics, on the other hand, are primarily related to safety constraints and are therefore handled separately in the subsequent safety layer design. Accordingly, the electrolyzer system is represented in the following plant-output form:
\begin{equation}\label{E12}
\left\{
\begin{aligned}
\dot x_T &= f(x_T,u) \\
y &= g(x_T,u)
\end{aligned}
\right.,
\end{equation}
where $x_T=[T_{ele,1},\dots,T_{ele,N_{ele}}]^\top$ denotes the temperature state vector of all electrolyzers, $u=[i_{ele,1},\dots,i_{ele,N_{ele}}]^\top$ is the corresponding operating current input vector, and $y=[P_{ele,1},\dots,P_{ele,N_{ele}}]^\top$ is the corresponding power output vector. Specifically, $f(\cdot)$ is determined by the thermal model in \eqref{E5}-\eqref{E7}, and $g(\cdot)$ is determined by the electrochemical power relation in \eqref{E1}-\eqref{E4}. For any fixed $u$, the temperature subsystem admits a unique steady state $\hat{x}_T(u)$ satisfying $0=f(\hat{x}_T(u),u)$. Accordingly, the steady state input-output map can be defined as
\begin{equation}\label{E13}
    h(u):=g(\hat{x}_T(u),u).
\end{equation}

For the power tracking objective in (P1), define the instantaneous power mismatch as
\begin{equation}\label{E14}
\Phi(y,t)=\frac{1}{2}\left(\sum_{i=1}^{N_{ele}} y_i-P_{wind}(t)\right)^2,
\end{equation}
where $y_i$ denotes the power consumption of the $i$-th electrolyzer. 
For real-time implementation, replacing the plant output $y$ by the steady state map $h(u)$ yields the time-varying surrogate objective
\begin{equation}\label{E15}
\Psi(u,t):=\Phi(h(u),t)
=
\frac{1}{2}\left(\sum_{i=1}^{N_{ele}} h_i(u_i)-P_{wind}(t)\right)^2.
\end{equation}

Therefore, the power tracking objective in (P1) is approximated online by reducing the surrogate objective $\Psi(u,t)$ with respect to $u$ at each time instant. Since the available wind power is time-varying, the above surrogate problem is a family of instantaneous optimization problems parameterized by $t$.
This approximation preserves the instantaneous power tracking objective while avoiding the need to solve a multi-step dynamic optimization problem online.
For a fixed current input $u_i$, the steady state temperature $\hat{T}_{ele,i}(u_i)$ is determined by the steady state condition of the thermal dynamics, i.e.,
\begin{equation}\label{E16}
0=(\rho_{1,i}+\rho_{2,i}\hat{T}_{ele,i})u_i^2-(\hat{T}_{ele,i}-T_a)/{R_{th,i}},
\end{equation}
which gives
\begin{equation}\label{E17}
\hat{T}_{ele,i}(u_i)=
(T_a+R_{th,i}\rho_{1,i}u_i^2)/
({1-R_{th,i}\rho_{2,i}u_i^2}).
\end{equation}

Substituting $\hat{T}_{ele,i}(u_i)$ into the electrolyzer power model yields the steady state input-output map
\begin{equation}\label{E18}
h_i(u_i)=
\left[
u_{rev,i}+
\left(\rho_{1,i}+\rho_{2,i}\hat{T}_{ele,i}(u_i)\right)u_i
\right]u_i .
\end{equation}

Based on the surrogate objective \eqref{E15}, a feedback optimization law is introduced to generate an online input update direction for reducing the instantaneous power mismatch. The corresponding dynamics are given by
\begin{equation}\label{E19}
    \dot u = -\varepsilon \nabla_u \Psi(u,t),
\end{equation}
where $\varepsilon$ denotes the controller gain. For sampled data implementation with sampling period $\Delta t$, the desired input generated by the feedback optimization layer is computed as
\begin{equation}\label{E20}
u_{des}(t+1)=u(t)-\varepsilon \Delta t \, \nabla_u \Psi\big(u(t),t\big),
\end{equation}
which serves as the nominal input update generated by the feedback optimization layer and is then processed by the projection-based safety layer in the following subsections.

\subsection{Safety constraints handling via CBF}\label{S3-2}
Since the HTO constraint in \eqref{E11g} is governed by internal states such as temperature and hydrogen impurity accumulation, rather than solely by the instantaneous control input, directly enforcing this constraint in the projection-based safety layer may easily render the optimization infeasible, especially when the system is already close to the safety boundary. Therefore, an additional mechanism is needed to explicitly characterize how the current input affects the subsequent HTO evolution and to transform the state safety requirement into a form suitable for real-time optimization. To this end, CBF theory is introduced to convert the HTO safety constraint from a state dependent condition into an input admissibility condition that can be directly incorporated into the safety layer \citep{P27,P28}. As the proposed safety layer is implemented in discrete time, the CBF conditions are formulated in their discrete-time forms in this paper.

For the $i$-th electrolyzer, let $x_{HTO,i}$ denote the internal state vector associated with the HTO dynamics, including the temperature and hydrogen accumulation states involved in Section \ref{S2}. The safe set associated with the HTO constraint is defined as
\begin{equation}\label{E21}
\mathcal{C}_{HTO,i} := \{x_{HTO,i} \mid h_{HTO,i}(x_{HTO,i}) \ge 0\},
\end{equation}
where
\begin{equation}\label{E22}
h_{HTO,i}(x_{HTO,i}) := HTO_{max} - HTO_i(x_{HTO,i}).
\end{equation}

Then, based on the discrete-time CBF condition in \citep{P29}, if the initial state satisfies $x_{HTO,i}(0)\in\mathcal{C}_{HTO,i}$, the safe set remains forward invariant, i.e., the HTO state stays within the safe region at all subsequent sampling instants, provided that
\begin{equation}\label{E23}
{h_{HTO,i}}\left( {x_{HTO,i}\left( {t + 1} \right)} \right) \ge \left( {1 - {\alpha}} \right){h_{HTO,i}}\left( {x_{HTO,i}\left( t \right)} \right),
\end{equation}
where ${\alpha} \in \left( {0,1} \right]$ is a design parameter. Substituting \eqref{E22} into \eqref{E23} yields
\begin{equation}\label{E24}
HTO_i\left(x_{HTO,i}(t+1)\right)
\le (1-\alpha)\,HTO_i\left(x_{HTO,i}(t)\right)+\alpha\,HTO_{max}.
\end{equation}

Thus, the HTO safety requirement is enforced through a one-step sufficient condition on the state evolution. Therefore, if the initial state satisfies the HTO safety constraint and \eqref{E24} is enforced at each sampling instant, the HTO safe set remains forward invariant. By applying Euler discretization to the thermal and HTO dynamics in Section \ref{S2} and substituting the resulting sampled data model into \eqref{E24}, the following explicit constraint on the control input is obtained:
\begin{equation}\label{E25}
k_{1,i}(t)u_i^3-k_{2,i}(t)u_i^2+k_{3,i}(t)u_i+k_{4,i}(t) \ge 0,
\end{equation}
where $k_{1,i}(t)$, $k_{2,i}(t)$, $k_{3,i}(t)$, and $k_{4,i}(t)$ are time-varying coefficients determined by the system state at time $t$. The detailed derivation is provided in Appendix A.

\subsection{Projection-based safety layer}\label{S3-3}
The feedback optimization law in Section \ref{S3-1} generates the desired control input $u_{des}(t+1)$ to improve wind power utilization. However, since $u_{des}(t+1)$ is obtained without explicitly enforcing the operational and safety constraints, it may fall outside the admissible operating region of the electrolyzers.
Meanwhile, as shown in Section \ref{S3-2}, the HTO safety requirement can be reformulated as an explicit admissibility condition on the control input using discrete-time CBF theory.
Therefore, a projection-based safety layer is introduced to enforce these constraints and compute the final safe control input online.

Specifically, at each sampling instant, the safety layer determines the next step control input that is closest to $u_{des}(t+1)$ in the Euclidean sense while satisfying the total power limits, electrical operating limits, ramp rate limits, and the CBF-based HTO safety constraints. In this way, the control intention of feedback optimization layer can be preserved as much as possible, while real-time constraint satisfaction is guaranteed. The resulting optimization problem is formulated as follows:
\begin{equation}\label{E26}
\begin{aligned}
\text{(P2)}\quad
&u^*(t+1)=\arg\min_{u(t+1)}\  \frac{1}{2}\left\|u(t+1)-u_{des}(t+1)\right\|^2\\
&\text{s.t.}\quad
 \eqref{E11b}-\eqref{E11f},\ \eqref{E25}.
\end{aligned}
\end{equation}

The solution $u^*(t+1)$ obtained from (P2) is applied to the electrolyzers as the final safe control input. In this way, the proposed method separates nominal power allocation input generation from constraint enforcement: the feedback optimization layer computes $u_{des}(t+1)$, while the projection-based safety layer guarantees that the implemented input remains feasible and safe. This structure helps reduce the online computational burden and facilitates real-time implementation. The overall algorithmic framework of the proposed method is summarized in Algorithm 1.

To ensure online implementability, it is necessary to characterize conditions under which problem (P2) is feasible. Since the objective in (P2) is continuous, the main issue is whether the admissible set defined by constraints \eqref{E11b}--\eqref{E11f} and \eqref{E25} is nonempty.
For the $i$-th electrolyzer, define the interval induced by the electrical operating limits and the ramp rate constraint as
\begin{equation}\label{E27}
\mathcal{I}_i(t):=[\underline{u}_i(t),\bar{u}_i(t)],
\end{equation}
where $\underline{u}_i(t)$ and $\bar{u}_i(t)$ are the lower and upper bounds determined by the current limit, voltage limit, power limit, and ramp rate limit. Then, define the local admissible set associated with the CBF-based HTO safety condition as
\begin{equation}\label{E28}
\mathcal{U}_i(t):=
\left\{
u_i \in \mathcal{I}_i(t):
k_{1,i}(t)u_i^3-k_{2,i}(t)u_i^2+k_{3,i}(t)u_i+k_{4,i}(t)\ge 0
\right\}.
\end{equation}

Furthermore, define the minimum admissible input for each electrolyzer and the corresponding minimum reachable total power as follows:
\begin{equation}\label{E29}
u_i^{-}(t):=\min\{u_i: u_i\in \mathcal U_i(t)\},
\end{equation}
\begin{equation}\label{E30}
P_{min}^{reach}(t):=
\sum_{i=1}^{N_{ele}} P_{ele,i}\big(u_i^{-}(t),t\big).
\end{equation}
\textbf{Proposition 1.}
At time $t$, problem (P2) is feasible if the following conditions hold:

$\bullet$ for each electrolyzer $i$, the local admissible set $\mathcal{U}_i(t)$ is nonempty;

$\bullet$ the available wind power satisfies
\begin{equation}\label{E31}
P_{min}^{reach}(t)\le P_{wind}(t).
\end{equation}
\textbf{Proof.}
The proof of Proposition 1 is listed in Appendix B.

Proposition 1 provides an explicit feasibility condition for the projection-based safety layer at each sampling instant. Based on this result, the properties of the overall framework are further discussed in the next section.

\begin{algorithm}[t]
\caption{Online coordinated operation of the proposed FO-safety layer framework}
\label{alg1}
\begin{algorithmic}[1]
\STATE \textbf{Input:} Current system state $x(t)$, previous control input $u(t)$, available \\wind power $P_{\mathrm{wind}}(t)$
\STATE \textbf{Output:} Safe control input $u(t+1)$
\WHILE{the system is in operation}
    \STATE Compute the FO-based input update direction $\dot{u}(t)$ according to \eqref{E19}
    \STATE Compute the desired input $u_{\mathrm{des}}(t+1)$ according to \eqref{E20}
    \IF{the condition in Proposition 1 is satisfied}
        \STATE Solve problem (P2) to obtain $u(t+1)$
    \ELSE
        \STATE Relax constraint \eqref{E11b} and solve the modified (P2)
    \ENDIF
    \STATE Apply $u(t+1)$ to the system dynamics and update the system state
    \STATE Set $t \gets t+1$
\ENDWHILE
\end{algorithmic}
\end{algorithm}
\section{Analysis of the feedback optimization layer and safety guarantee}\label{S4}
This section analyzes the proposed method from three aspects, namely, the discrete-time regulation property of the feedback optimization layer, the safety guarantee induced by the discrete-time CBF condition, and the practical closed-loop property of the overall implementation.
\subsection{Discrete-time regulation property of the feedback optimization layer}
Recall from Section 3.1 that the steady state temperature is given by \eqref{E17}, and the corresponding steady state power map is defined in \eqref{E18}. 

\noindent\textbf{Lemma 1.}
For each electrolyzer, the steady state power map $h_i(u_i)$ in \eqref{E18} is well defined and continuously differentiable over the admissible operating range. Moreover, under the practical condition $\rho_{1,i}+\rho_{2,i}T_a>0$, $h_i(u_i)$ is strictly increasing with respect to $u_i$.

\noindent\textbf{Proof.}
Substituting \eqref{E17} into \eqref{E18} yields
\begin{equation}\label{E32}
h_i(u_i)=u_{rev,i}u_i+
((\rho_{1,i}+\rho_{2,i}T_a)u_i^2)/({1-R_{th,i}\rho_{2,i}u_i^2}).
\end{equation}

Since $\rho_{2,i}\le 0$, the denominator in \eqref{E32} remains strictly positive for all $u_i\ge 0$, and therefore $h_i(u_i)$ is well defined and continuously differentiable. Differentiating \eqref{E32} gives
\begin{equation}\label{E33}
\frac{d h_i}{d u_i}
=
u_{rev,i}
+
\frac{2(\rho_{1,i}+\rho_{2,i}T_a)u_i}{(1-R_{th,i}\rho_{2,i}u_i^2)^2}.
\end{equation}

Under $\rho_{1,i}+\rho_{2,i}T_a>0$, one has $\frac{d h_i}{d u_i}>0$, which implies that $h_i(u_i)$ is strictly increasing over the admissible range. \hfill$\blacksquare$

For the discrete-time implementation in \eqref{E20}, define the power mismatch as
\begin{equation}\label{E34}
e(t):=\sum_{i=1}^{N_{ele}} h_i(u_i(t)) - P_{wind}(t).
\end{equation}

According to \eqref{E15}, one has
\begin{equation}\label{E35}
\nabla_u \Psi(u(t),t)=\nabla h(u(t))^\top e(t),
\end{equation}
where $\nabla h(u(t))=\left[\frac{dh_1}{du_1},\dots,\frac{dh_{N_{ele}}}{du_{N_{ele}}}\right]$.

\noindent\textbf{Proposition 2.}
Assume that the current input and the desired input generated by \eqref{E20} remain in a compact admissible region, and that the wind power variation is bounded, i.e., there exists a constant $\bar{\omega}\ge 0$ such that $|P_{wind}(t+1)-P_{wind}(t)|\le \bar{\omega}$. Then, if the gain $\varepsilon$ is chosen sufficiently small, the power mismatch associated with the desired input admits a one-step upper bound consisting of a contraction term and the bounded wind power variation.

\noindent\textbf{Proof.}
By Lemma 1, each $h_i(u_i)$ is continuously differentiable and strictly increasing over the admissible range. Hence, the map $\sum_{i=1}^{N_{ele}} h_i(u_i)$ is continuously differentiable, and its gradient is bounded and positive on any compact admissible region. Applying the mean value theorem to the update from $u(t)$ to $u_{des}(t+1)$ gives
\begin{align}
\sum_{i=1}^{N_{ele}} h_i(u_{des,i}(t+1))
&=
\sum_{i=1}^{N_{ele}} h_i(u_i(t))
-
\varepsilon \Delta t\, \beta_t\, e(t), \label{E36}
\end{align}
where $\beta_t$ is a positive bounded scalar determined by the gradients of the map at intermediate points. Therefore, for sufficiently small $\varepsilon$, there exists a constant $q\in(0,1)$ such that
\begin{equation}\label{E37}
\left|
\sum_{i=1}^{N_{ele}} h_i(u_{des,i}(t+1)) - P_{wind}(t)
\right|
\le q |e(t)|.
\end{equation}

Combining \eqref{E37} with the assumption that the wind power variation is bounded yields
\begin{equation}\label{E38}
\left|
\sum_{i=1}^{N_{ele}} h_i(u_{des,i}(t+1)) - P_{wind}(t+1)
\right|
\le q |e(t)|+\bar{\omega}.
\end{equation}

Hence, the power mismatch associated with the desired input admits the one-step upper bound. \hfill$\blacksquare$

\subsection{Safety guarantee induced by the discrete-time CBF condition}

The HTO safety constraint is enforced through the discrete-time CBF condition in \eqref{E23}. The following result states the corresponding forward invariance property.

\noindent\textbf{Proposition 3.}
Assume that the initial HTO-related state of each electrolyzer satisfies $x_{HTO,i}(0)\in \mathcal{C}_{HTO,i}$, $i=1,\dots,N_{ele}$. If the implemented input satisfies the discrete-time CBF condition \eqref{E23} at every sampling instant, then the safe set $\mathcal{C}_{HTO,i}$ is forward invariant for each electrolyzer.

\noindent\textbf{Proof.}
Since $x_{HTO,i}(0)\in \mathcal{C}_{HTO,i}$, one has $h_{HTO,i}(x_{HTO,i}(0))\ge 0$. Suppose that $h_{HTO,i}(x_{HTO,i}(t))\ge 0$ at some sampling instant $t$. Then, by \eqref{E23},
\begin{equation}\label{E39}
h_{HTO,i}(x_{HTO,i}(t+1))
\ge
(1-\alpha)h_{HTO,i}(x_{HTO,i}(t))
\ge 0.
\end{equation}

Thus, by induction, $h_{HTO,i}(x_{HTO,i}(t))\ge 0$ for all $t\ge 0$, and therefore $x_{HTO,i}(t)\in \mathcal{C}_{HTO,i}$ for all $t\ge 0$. \hfill$\blacksquare$

\subsection{Practical closed-loop property of the overall layered framework}

The above results can be combined with the feasibility result in Proposition 1 to establish a practical closed-loop result for the overall layered framework.

\noindent\textbf{Theorem 1.}
Suppose that the conditions of Propositions 1-3 hold, and that there exists a constant $\bar{e}_f\ge 0$ satisfying $\|u(t+1)-u_{des}(t+1)\|\le \bar{e}_f$. Then, under the proposed layered implementation:

$\bullet$ problem (P2) admits at least one optimal solution at each sampling instant;

$\bullet$ the implemented input is admissible and the HTO safe set remains forward invariant;

$\bullet$ the power mismatch associated with the implemented input is uniformly ultimately bounded.

\noindent\textbf{Proof.}
Considering the actual power mismatch associated with the implemented input at time $t+1$, denoted by $e(t+1)$, one obtains
\begin{equation}\label{E40}
\begin{aligned}
|e(t+1)|
\le{}\;&
\left|
\sum_{i=1}^{N_{ele}} h_i\bigl(u_{des,i}(t+1)\bigr)
- P_{wind}(t+1)
\right|
\\
&+
\left|
\sum_{i=1}^{N_{ele}} h_i\bigl(u_i(t+1)\bigr)
-
\sum_{i=1}^{N_{ele}} h_i\bigl(u_{des,i}(t+1)\bigr)
\right|.
\end{aligned}
\end{equation}

By Proposition 2, the first term on the right-hand side satisfies the one-step bound in \eqref{E38}. Since each $h_i$ is continuously differentiable over the admissible region, the map $\sum_{i=1}^{N_{ele}} h_i(u_i)$ is Lipschitz continuous on the region. Therefore, there exists a constant $L_h>0$ such that
\begin{equation}\label{E41}
\left|
\sum_{i=1}^{N_{ele}} h_i(u_i(t+1))
-
\sum_{i=1}^{N_{ele}} h_i(u_{des,i}(t+1))
\right|
\le
L_h \|u(t+1)-u_{des}(t+1)\|.
\end{equation}

By the assumed boundedness of the safety layer deviation, one has
\begin{equation}\label{E42}
\left|
\sum_{i=1}^{N_{ele}} h_i(u_i(t+1))
-
\sum_{i=1}^{N_{ele}} h_i(u_{des,i}(t+1))
\right|
\le
L_h \bar{e}_f.
\end{equation}

Combining \eqref{E38}, \eqref{E40}, and \eqref{E42} yields
\begin{equation}\label{E43}
|e(t+1)|
\le
q |e(t)| + \bar{\omega} + L_h \bar{e}_f.
\end{equation}

By recursively applying \eqref{E43}, one obtains
\begin{equation}\label{E44}
|e(t)|
\le
q^t |e(0)|
+
\frac{1-q^t}{1-q}\big(\bar{\omega}+L_h \bar{e}_f\big).
\end{equation}

Hence, the actual power mismatch is uniformly ultimately bounded. \hfill$\blacksquare$

The above analysis shows that the proposed framework possesses three key properties: the FO layer practically regulates the power mismatch in discrete time, the discrete-time CBF condition guarantees forward invariance of the HTO safe set, and the overall layered implementation preserves admissibility and safety while maintaining practical closed-loop performance.
\section{Numerical results}\label{S5}
\subsection{Experimental setup and representative day construction}\label{S5-1}
\begin{figure}[t]
	\centering
	\includegraphics[scale=0.32]{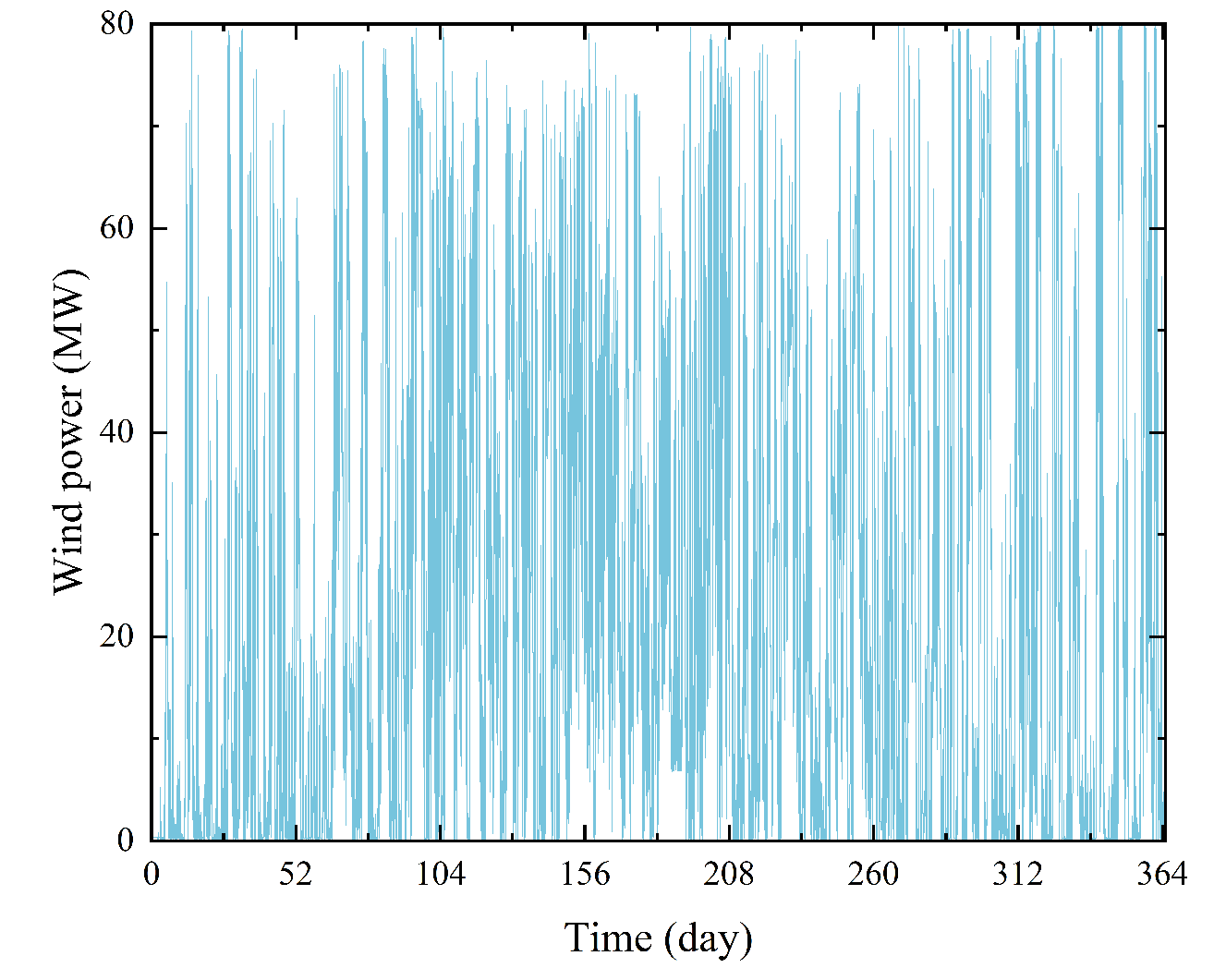}
	\caption{Annual wind power.}
	\label{Figure 2}
\end{figure}

\begin{figure*}[t]
	\centering  
	\includegraphics[scale=0.4]{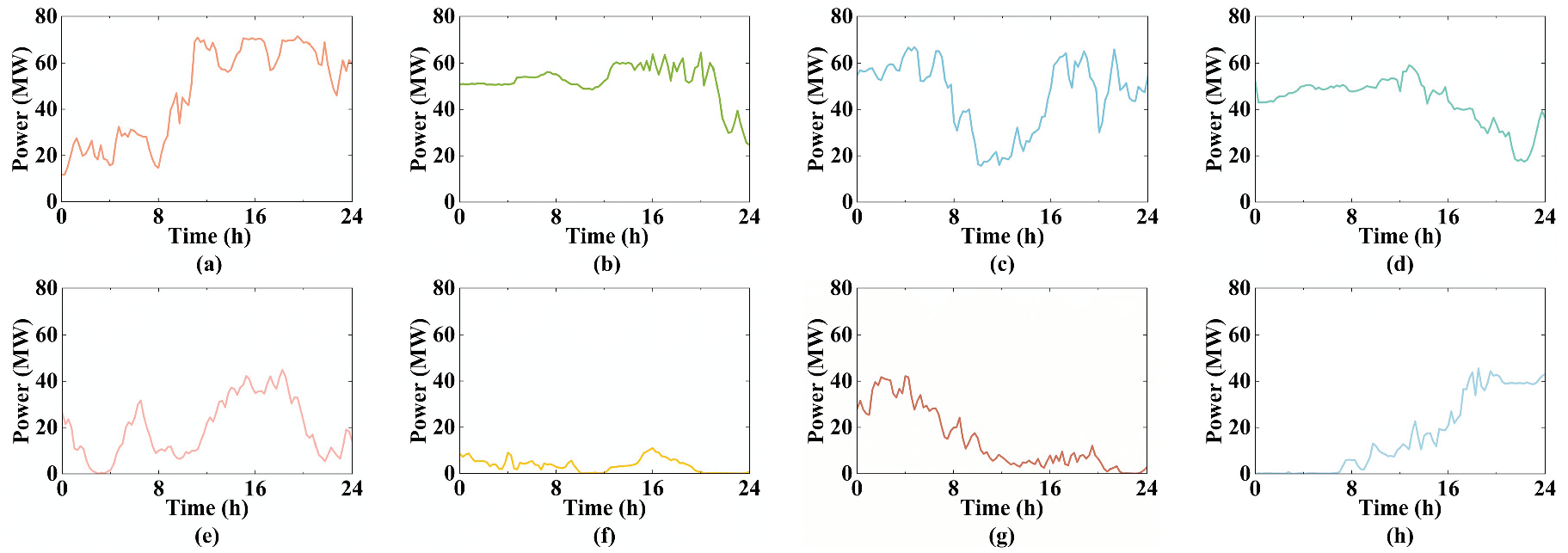}
	\caption{Wind power curves of the representative days.} 
	\label{Figure 3} 
\end{figure*}

To verify the effectiveness of the proposed method under realistic renewable power fluctuations, numerical simulations are performed using the full year wind power generation data of a wind farm in China in 2019 \citep{P30}. For numerical testing, the wind power profile is scaled to match the power range of the studied electrolyzer system, namely, the total rated power determined by the number of electrolyzers and the rated power of each unit. The corresponding annual wind power profile is presented in Fig. \ref{Figure 2}. To capture the diversity of wind power dynamics throughout the year and construct representative operating scenarios for numerical evaluation, K-means clustering is adopted to extract representative days from the annual dataset \citep{P31}, as shown in Fig. \ref{Figure 3}. Based on these representative days, the proposed method is evaluated with respect to tracking performance and operational safety under different renewable generation conditions. The main electrolyzer parameters used in the simulations are listed in Table \ref{tab2}.
\begin{table}[t]
\centering
\caption{Parameters of electrolyzers.}
\label{tab2}
\resizebox{0.98\columnwidth}{!}{%
\begin{tabular}{cc}
\toprule
Parameter&Value\\
\midrule
Number of electrolytic cells $N_{cell}$&45\\
Voltage limit of a single cell $u_{limit}$&$2.1V$\\
Electrochemical parameter $\rho_1$&$3.11$\\
Electrochemical parameter $\rho_2$&$-0.025$\\
Thermal resistance $R_{th}$&$0.054K/W$\\
Thermal capacity $C_{th}$&$15000J/K$\\
Ambient temperature $T_a$&$25^\circ\mathrm{C}$\\
Controller gain $\varepsilon$&$0.00001$\\
sampling period $\Delta t$ &$1s$\\
CBF parameter $\alpha$&$0.8$\\
HTO safety limit $HTO_{max}$&$0.02$\\
Initial temperature &$[25.0, 30.0, 40.0, 60.0]^\circ\mathrm{C}$\\
\bottomrule
\end{tabular}%
}
\end{table}

\subsection{Performance, safety, and real-time applicability validation}\label{S5-2}
Based on the representative days constructed in Section \ref{S5-1}, this subsection validates the performance and safety of the proposed method under typical renewable generation conditions. Simulations are first conducted on a system consisting of four electrolyzers to examine whether the proposed method can achieve effective power tracking while maintaining safe operation. In addition, the corresponding computational results are presented to evaluate its real-time implementation capability.

\begin{figure*}[t]
	\centering  
	\includegraphics[scale=0.4]{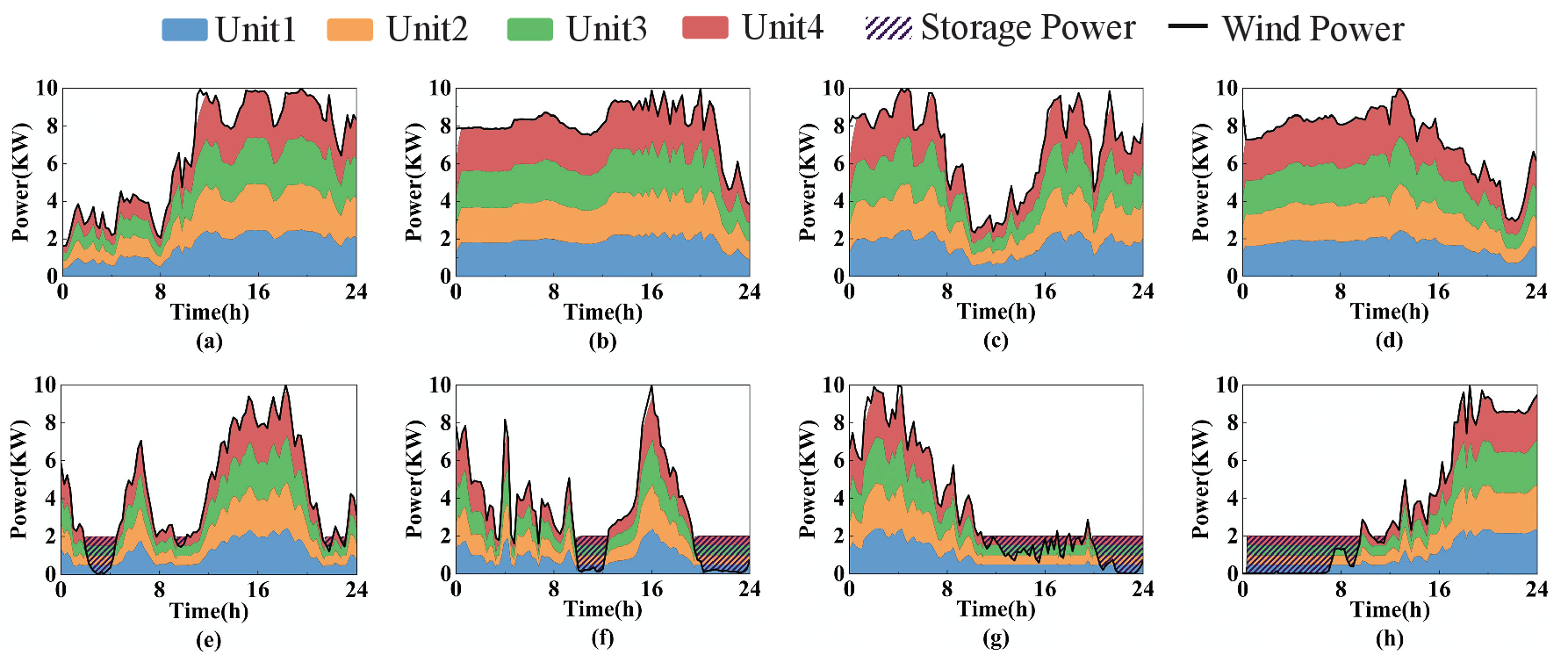}
	\caption{Power allocation and utilization results of the four electrolyzer system under the representative day scenarios.} 
	\label{Figure 4} 
\end{figure*}

Fig. \ref{Figure 4} presents the power utilization results of the proposed method for the four electrolyzer system. The total power consumption of the electrolyzer cluster closely follows the available wind power profiles of the representative day scenarios. In Fig. \ref{Figure 4}(a)-(d), wind generation alone is sufficient to sustain safe operation, such that the power demand is entirely supplied by wind power and the four electrolyzers exhibit coordinated load sharing behavior. In Fig. \ref{Figure 4}(e)-(h), however, the wind power remains at a relatively low level for an extended period, causing the electrolyzers to stay in the allowable low power operating region for a long time. Under these conditions, wind power alone is no longer sufficient to satisfy the safety requirement, and storage support becomes necessary to provide the additional power needed for maintaining safe operation.

\begin{figure*}[t]
	\centering  
	\includegraphics[scale=0.4]{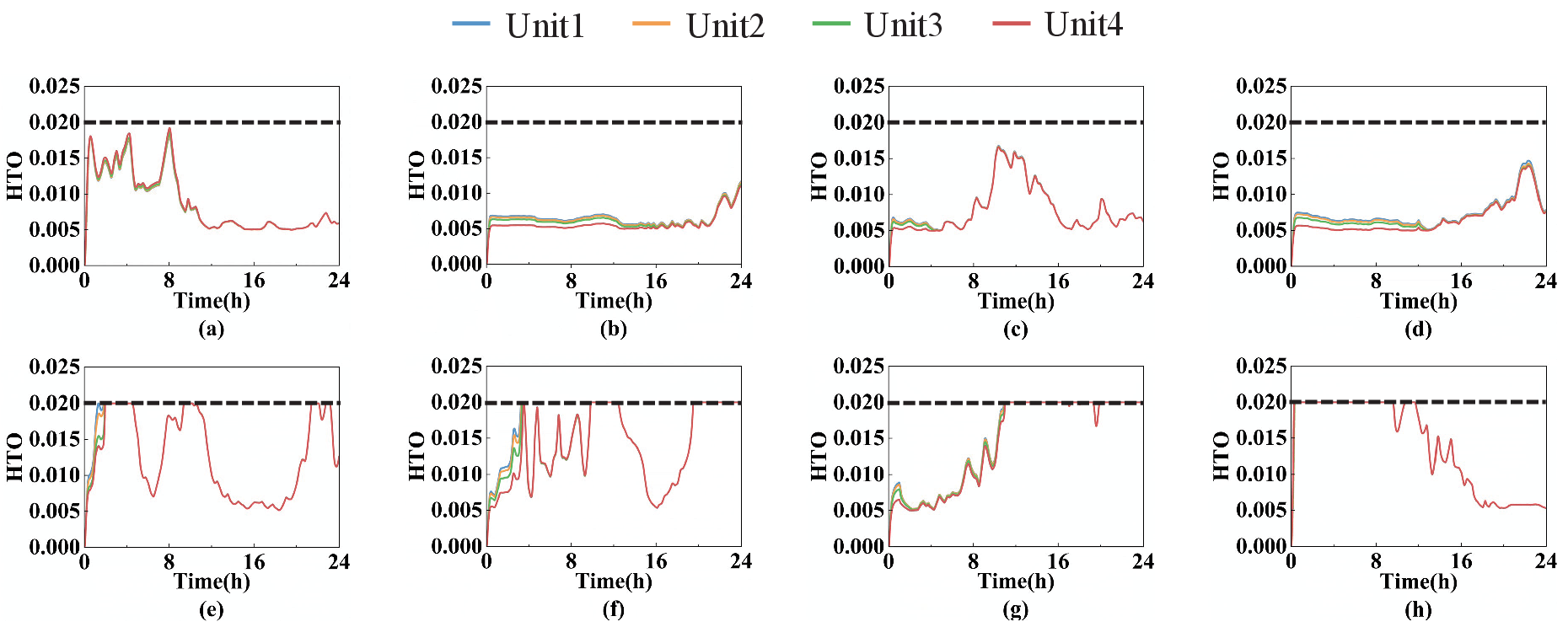}
	\caption{HTO trajectories of the four electrolyzer system under the representative day scenarios.} 
	\label{Figure 5} 
\end{figure*}

The corresponding HTO trajectories in Fig. \ref{Figure 5} further verify the safety preserving property of the proposed method. Throughout the entire simulation horizon, the HTO levels of all electrolyzers remain below the prescribed safety limit, indicating that safe operation is consistently maintained under fluctuating renewable power input. In scenarios where wind power alone cannot sustain the required safe operating condition, storage support provides the additional power needed to keep the system within the admissible safety region. This confirms that the role of storage in the proposed method is inherently safety oriented.

\begin{figure*}[t]
	\centering  
	\includegraphics[scale=0.4]{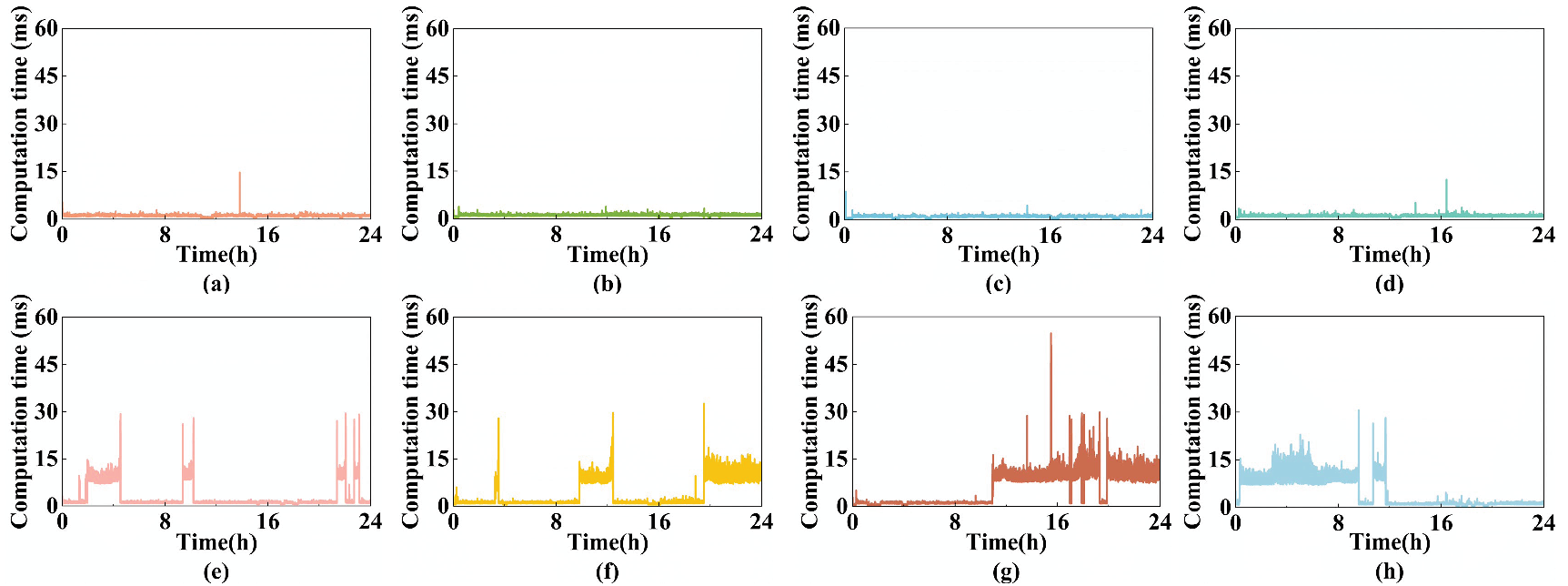}
	\caption{Online solve time trajectories under the representative day scenarios.} 
	\label{Figure 6} 
\end{figure*}

To further evaluate the real-time computational performance, Fig. \ref{Figure 6} illustrates the solve time trajectories under the representative day scenarios, and Table \ref{tab3} summarizes the associated statistics. It can be seen that the computational burden increases when storage support is introduced to preserve safe operation. In addition, more pronounced wind power fluctuations generally lead to higher solve times, while relatively smooth wind power profiles result in lower computational cost. For all representative day scenarios, however, the maximum solve time remains below 55 ms, which is well below the sampling interval $\Delta t$. This further confirms the practical implementability of the proposed method.
\begin{table}[t]
\centering
\caption{Solve time statistics under the representative day scenarios.}
\label{tab3}
\begin{tabular}{cccc}
\toprule
Scenario & Average (ms) & 95th percentile (ms) & Maximum (ms) \\
\midrule
Day (a) & 1.068 & 1.299 & 14.642 \\
Day (b) & 1.080 & 1.321 & 3.849 \\
Day (c) & 1.027 & 1.280 & 8.775 \\
Day (d) & 1.081 & 1.312 & 12.440 \\
Day (e) & 2.773 & 9.886 & 29.500 \\
Day (f) & 3.678 & 10.674 & 32.616 \\
Day (g) & 5.761 & 12.149 & 54.730 \\
Day (h) & 4.801 & 11.652 & 30.625 \\
\midrule
Mean & 2.659 & 6.197 & 23.397 \\
\bottomrule
\end{tabular}
\end{table}

\subsection{Scalability validation}
The results in Section \ref{S5-2} verify the effectiveness of the proposed method for the four electrolyzer case. To further examine its applicability in larger scale systems, this subsection extends the study to a ten electrolyzer configuration. The objective is to evaluate whether the proposed method can still achieve effective power allocation and maintain safe operation as the system scale increases.

\begin{figure*}[t]
	\centering  
	\includegraphics[scale=0.4]{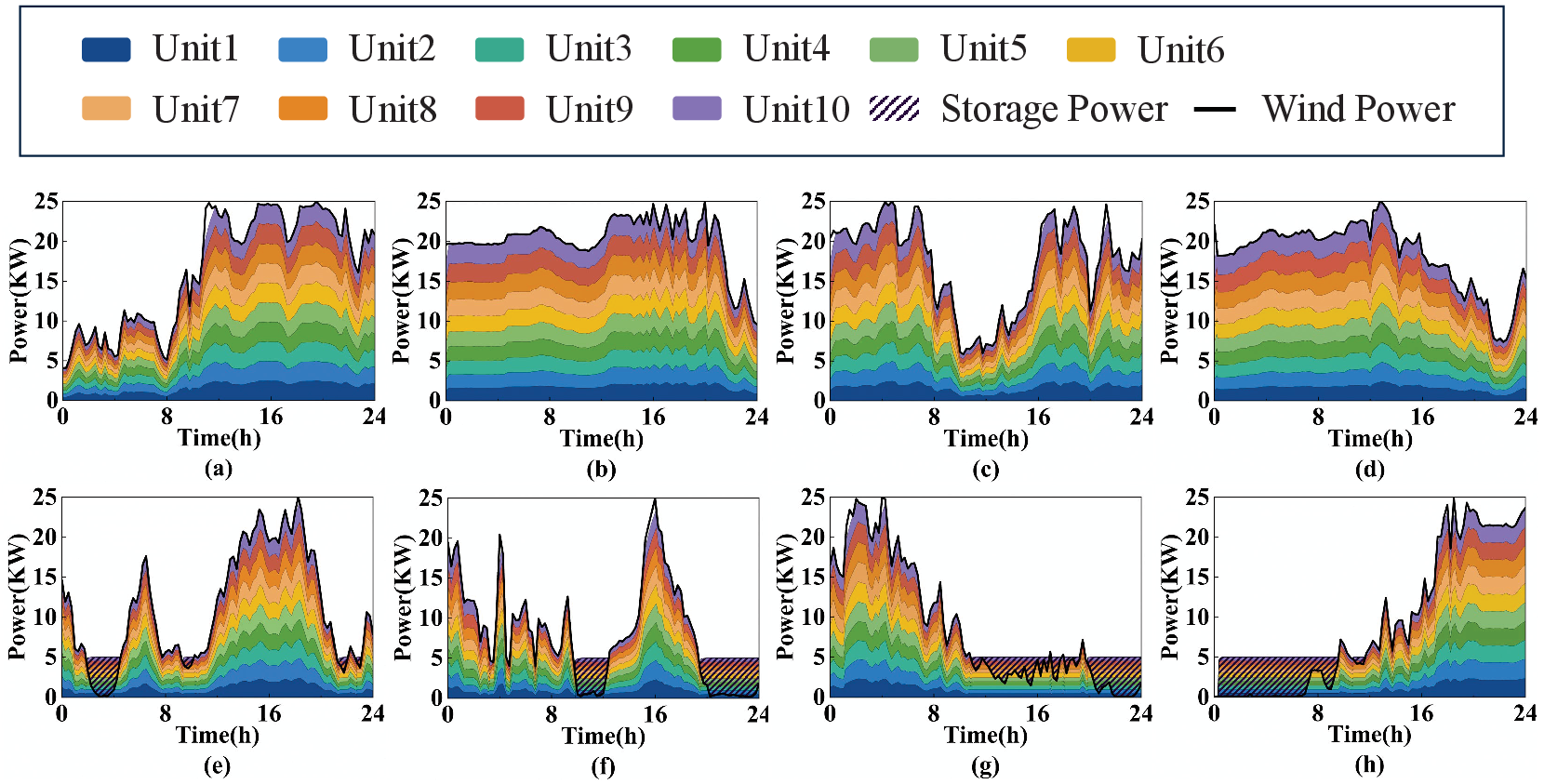}
	\caption{Power allocation and utilization results of the ten electrolyzer system under the representative day scenarios.} 
	\label{Figure 7} 
\end{figure*}

\begin{figure*}[t]
	\centering  
	\includegraphics[scale=0.4]{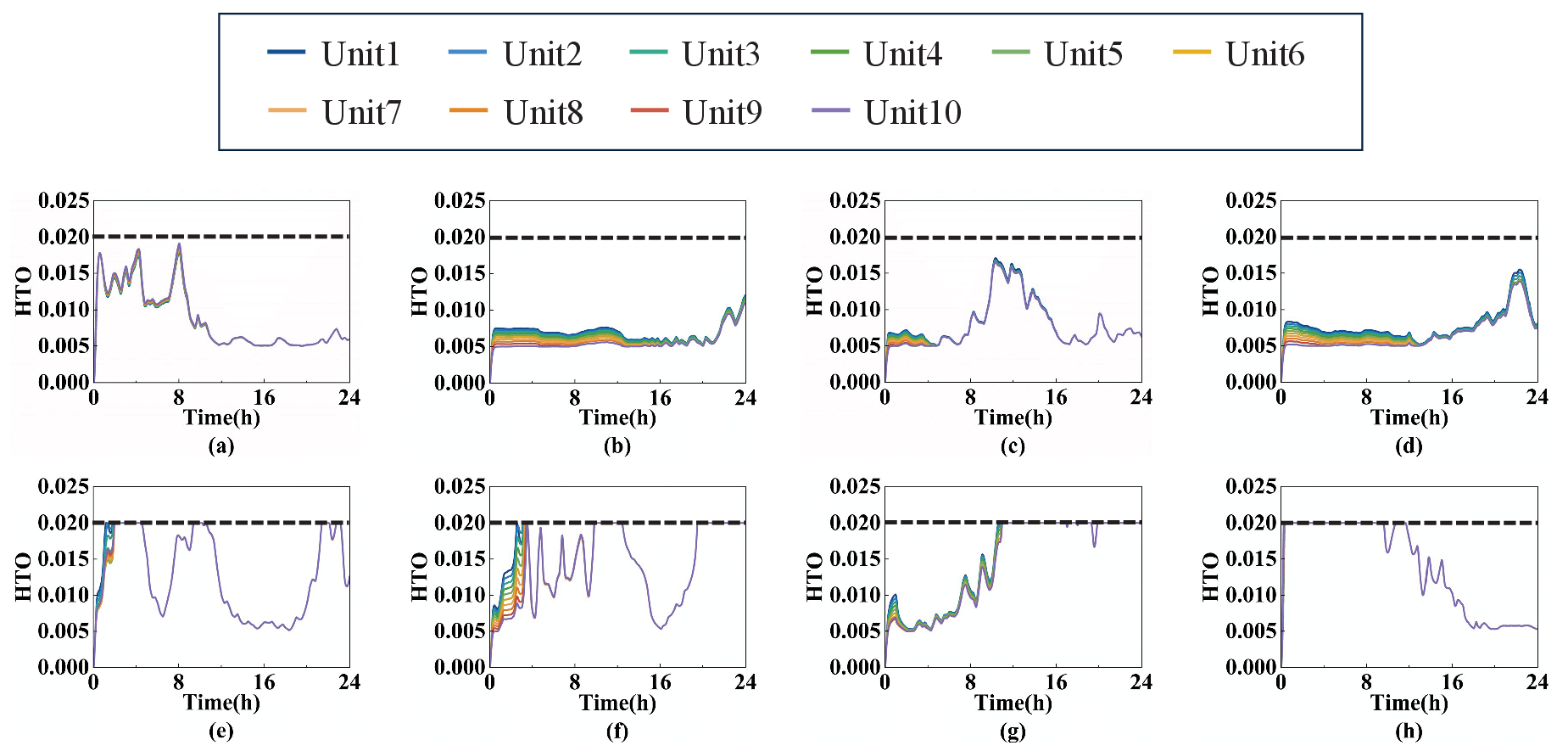}
	\caption{HTO trajectories of the ten electrolyzer system under the representative day scenarios.} 
	\label{Figure 8} 
\end{figure*}

Fig. \ref{Figure 7} and Fig. \ref{Figure 8} present the power utilization results and the corresponding HTO trajectories for the ten electrolyzer system under the representative day scenarios. It can be observed that, even when the system scale is increased, the proposed method still achieves coordinated power allocation among the electrolyzers while maintaining safe operation throughout the entire simulation horizon. Similar to the four electrolyzer case, wind power alone is sufficient to support safe operation in some scenarios, whereas storage support becomes necessary in low wind power conditions to provide the additional power required for safety preservation. Meanwhile, the HTO levels of all electrolyzers remain below the prescribed safety limit in all representative day scenarios, confirming that the safety preserving property of the proposed method is retained under the larger scale configuration. These results demonstrate that the proposed method can effectively scale to a larger electrolyzer cluster without losing its power allocation capability or safety guarantee.

\subsection{Sensitivity analysis of key design parameters}
To further examine the influence of key design parameters on the proposed method, this subsection presents a sensitivity study with respect to the feedback optimization gain $\varepsilon$ and the CBF coefficient $\alpha$ in the four electrolyzer case. The four electrolyzer setting is adopted as the baseline testbed because it is sufficiently representative to capture the closed-loop interaction between the feedback optimization layer and the projection-based safety layer, while avoiding additional scaling effects introduced by larger cluster sizes. For the sensitivity with respect to $\varepsilon$, the energy utilization rate is chosen as the main performance index, since it directly reflects how effectively the electrolyzer cluster tracks the available renewable power over the entire simulation horizon. For the sensitivity with respect to $\alpha$, the storage energy usage is selected as the main indicator because the CBF coefficient primarily affects the conservativeness of the safety layer, which in turn influences the amount of storage support required to preserve safe operation during transient processes. The gain sensitivity is presented in terms of a gain factor relative to the nominal controller gain listed in Table \ref{tab2}. In addition, for different values of $\alpha$, the storage energy usage is computed for each representative day to evaluate the conservativeness of the safety layer.

\begin{figure}[t]
	\centering  
	\includegraphics[scale=0.32]{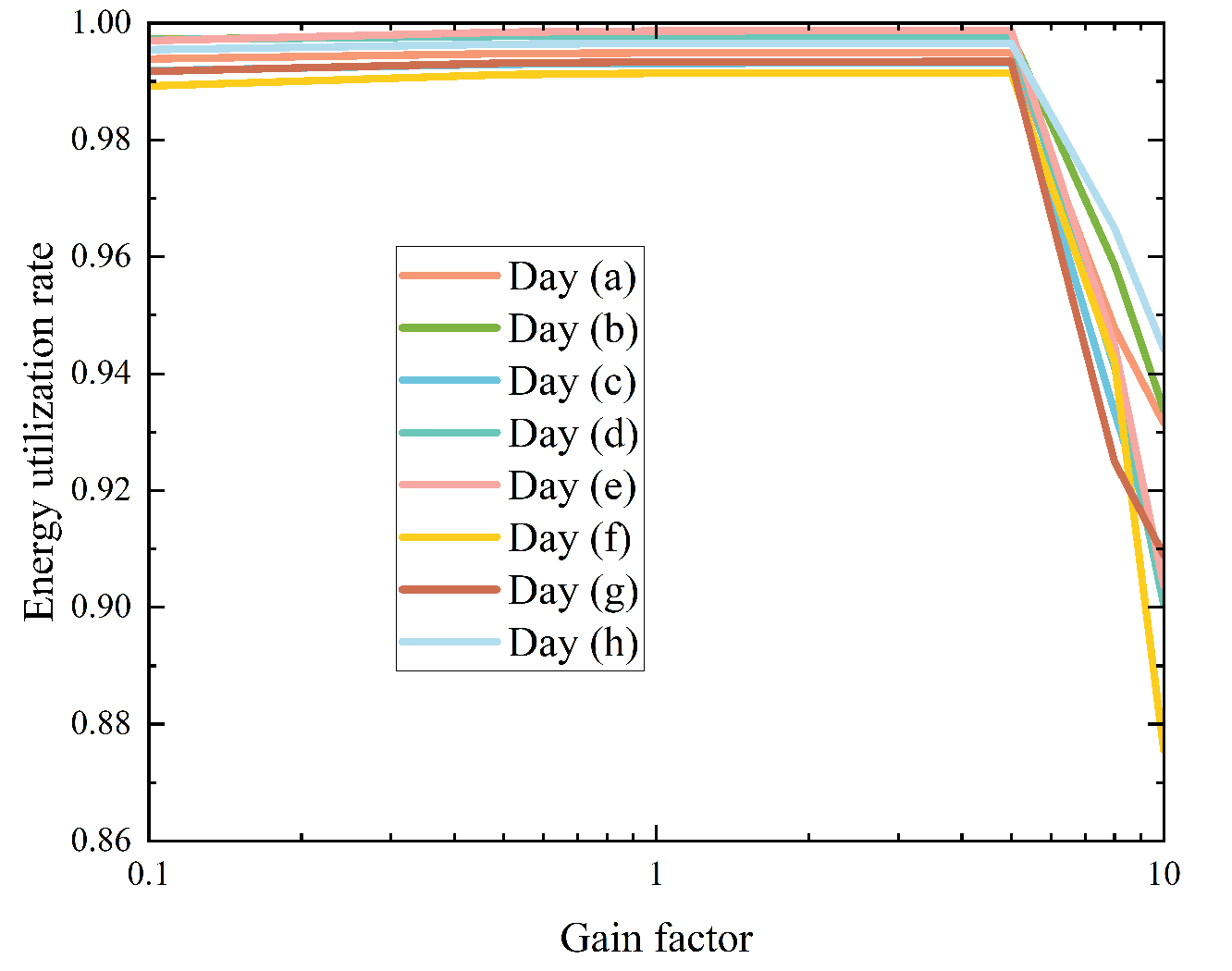}
	\caption{Sensitivity of the energy utilization rate to the feedback optimization gain factor.} 
	\label{Figure 9} 
\end{figure}

Fig. \ref{Figure 9} shows the sensitivity of the energy utilization rate to the feedback optimization gain. When the gain is excessively small, the control input is updated conservatively, resulting in a relatively slow tracking response and thus a lower utilization of the available renewable power. As the gain increases from this small gain region, the tracking speed improves and the energy utilization rate rises gradually. However, when the gain becomes overly large, the closed-loop response becomes more oscillatory, which deteriorates the power tracking performance and leads to a pronounced drop in energy utilization. It is worth noting that the proposed method maintains a high and nearly unchanged utilization rate over a relatively broad intermediate range of gain factors. This indicates that the closed-loop performance is not overly sensitive to the exact tuning of $\varepsilon$, thereby demonstrating a desirable degree of robustness with respect to the controller gain.

\begin{figure}[t]
	\centering  
	\includegraphics[scale=0.32]{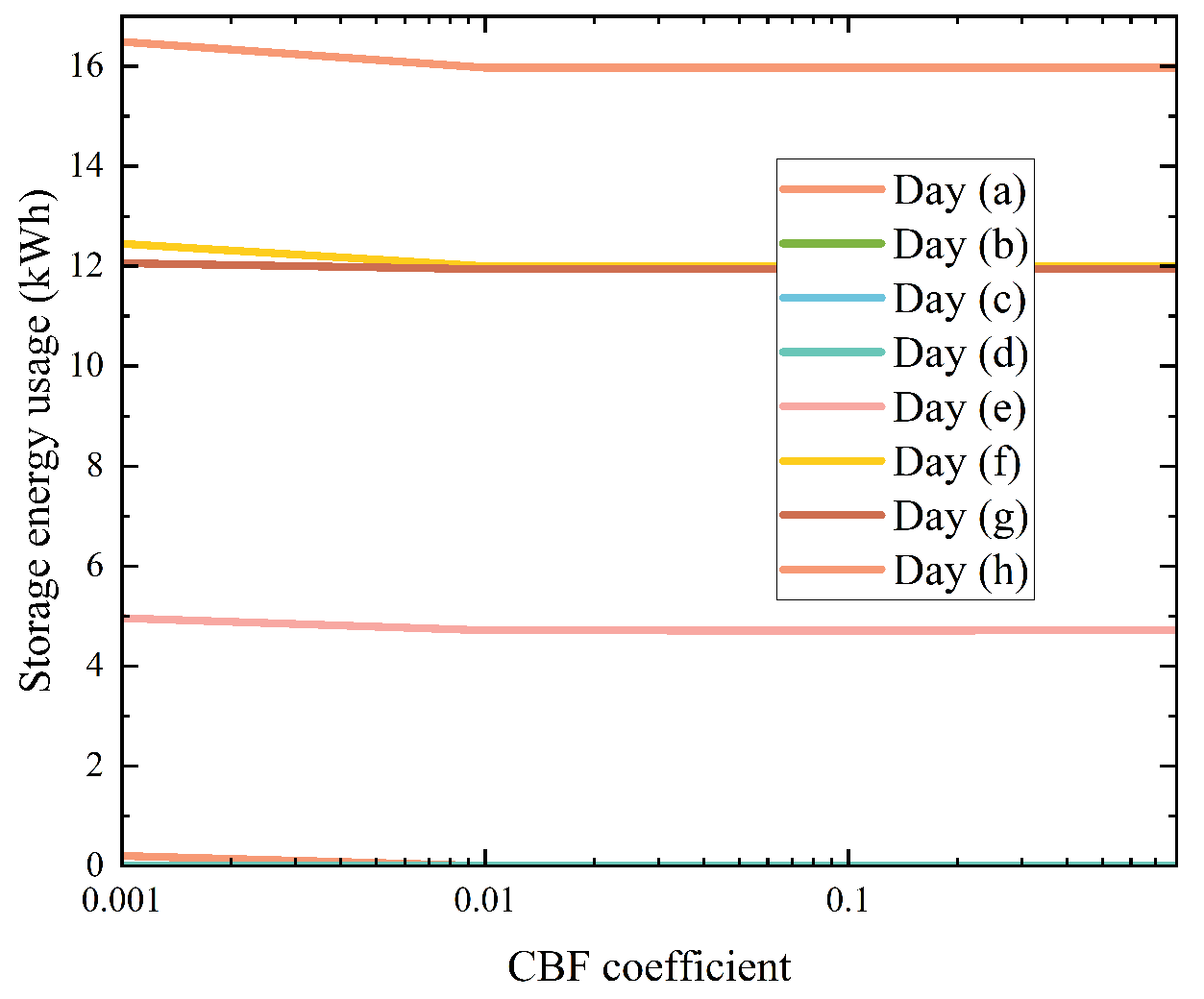}
	\caption{Sensitivity of the storage energy usage to the CBF coefficient.} 
	\label{Figure 10} 
\end{figure}

Fig. \ref{Figure 10} illustrates the influence of the CBF coefficient on the storage energy usage. When $\alpha$ is small, the discrete-time CBF condition remains relatively strict even when the operating point is still away from the safety boundary. As a result, the safety layer intervenes more conservatively, which increases the reliance on storage support during transient operation. By contrast, increasing $\alpha$ relaxes the one-step safety requirement to an appropriate extent, thereby reducing unnecessary conservativeness and lowering the storage energy usage while still preserving safety satisfaction. Meanwhile, the numerical results show that a larger $\alpha$ can significantly reduce the average computation time of the online optimization. This is because a less conservative safety constraint typically decreases the frequency and intensity of projection corrections, making the corresponding optimization problem easier to solve. Therefore, a proper increase in $\alpha$ improves computational efficiency without violating the safety constraints, which is favorable for real-time implementation.
\section{Conclusions}\label{S6}
This paper presents a coordinated operation framework for off-grid wind powered multi-electrolyzer systems by integrating feedback optimization with a projection-based safety layer. Thermal dynamics are explicitly incorporated into the coordinated operation design, and the HTO safety constraints are rigorously handled through a discrete-time CBF theory, thereby enabling real-time coordinated power allocation while preserving operational safety. Numerical studies based on real wind power data demonstrate that the proposed framework can achieve effective renewable power utilization while preserving safe operation under representative scenarios. In addition, the results confirm its real-time applicability and scalability, and the sensitivity analysis further verifies the robustness of the framework and provides guidance for parameter tuning in practical applications.

\appendix
\renewcommand{\theequation}{A.\arabic{equation}}
\setcounter{equation}{0}
\section*{Appendix A. Derivation of the CBF-based HTO constraint}\label{APP-A}

Using Euler discretization with sampling period $\Delta t$, the thermal and HTO-related dynamics under the normal operating mode can be written as follows:
\begin{equation}
T_{ele}(t+1) = T_{ele}(t)
- \frac{\Delta t}{R_{th}C_{th}}\bigl(T_{ele}(t)-T_a\bigr)
+ \frac{\Delta t}{C_{th}}
\bigl(\rho_1+\rho_2T_{ele}(t)\bigr)u^2 .
\end{equation}
\begin{equation}
{n_{an}}\left( {t + 1} \right) = {n_{an}}\left( t \right) + \left( {{{\dot n}_{cross}} - {n_{an}}\left( t \right){v_{lye}}/{V_{an}}/2} \right)\Delta t,
\end{equation}
\begin{equation}
n_{sep,l}(t+1)=n_{sep,l}(t)+\left(
n_{an}(t)\frac{v_{lye}}{2V_{an}}
-\frac{n_{sep,l}(t)}{\tau_{sep,l}}
\right)\Delta t .
\end{equation}
\begin{equation}
n_{sep,g}(t+1) = n_{sep,g}(t)
+ \frac{n_{sep,l}(t)}{\tau_{sep,l}}\,\Delta t 
- \frac{\eta_F N_{cell}\,n_{sep,g}(t)\,T_{ele}(t)\,R\,u}
{2 z_H F P V_{sep,g}}\,\Delta t .
\end{equation}
\begin{equation}
HTO(t) = \frac{T_{ele}(t)\,n_{sep,g}(t)\,R}{P\,V_{sep,g}},
\end{equation}
\begin{equation}
HTO\left( t + 1 \right) \le \left( {1 - \alpha } \right)HTO\left( t \right) + \alpha HT{O_{max }}.
\end{equation}

Substituting the above sampled data model into the discrete-time CBF condition and rearranging the terms with respect to $u$ yields
\begin{equation}
\small
\begin{aligned}
&\underbrace{
\frac{\Delta t}{C_{th}}
\left(\rho_1+\rho_2 T_{ele}(t)\right)
\frac{\eta_F N_{cell}\Delta t}{z_H F}
\frac{n_{sep,g}(t)\,T_{ele}(t)\,R}{2 P V_{sep,g}}
}_{k_1} u^3
\\[2mm]
&\quad
-\underbrace{
\frac{\Delta t}{C_{th}}
\left(\rho_1+\rho_2 T_{ele}(t)\right)
\left(
n_{sep,g}(t)+\frac{n_{sep,l}(t)\Delta t}{\tau_{sep,l}}
\right)
}_{k_2} u^2
\\[2mm]
&\quad
+\underbrace{
\left(
T_{ele}(t)-\frac{\Delta t}{R_{th}C_{th}}\left(T_{ele}(t)-T_a\right)
\right)
\frac{\eta_F N_{cell}\Delta t}{z_H F}
\frac{n_{sep,g}(t)\,T_{ele}(t)\,R}{2 P V_{sep,g}}
}_{k_3} u
\\[2mm]
&\quad
+\underbrace{
\left(
T_{ele}(t)n_{sep,g}(t)
+\alpha\left(
HTO_{max}\frac{P V_{sep,g}}{R}
-T_{ele}(t)n_{sep,g}(t)
\right)
\right)
}_{k_{4,1}}
\\
&\quad
-\underbrace{
\left(
T_{ele}(t)-\frac{\Delta t}{R_{th}C_{th}}\left(T_{ele}(t)-T_a\right)
\right)
\left(
n_{sep,g}(t)+\frac{n_{sep,l}(t)\Delta t}{\tau_{sep,l}}
\right)
}_{k_{4,2}}
\ge 0 .
\end{aligned}
\end{equation}

For notational simplicity in the main text, define
\begin{equation}
k_4 := k_{4,1}-k_{4,2},
\end{equation}

Then, the CBF-based HTO constraint can be written compactly as
\begin{equation}
k_1u^3-k_2u^2+k_3u+k_4\ge 0.
\end{equation}

\section*{Appendix B. Proof of Proposition 1}
\setcounter{equation}{0}
\renewcommand{\theequation}{B.\arabic{equation}}

For the $i$-th electrolyzer, the admissible current interval induced by the operational constraints is defined as
\begin{equation}
\mathcal I_i(t):=[\underline{u}_i(t),\bar{u}_i(t)],
\label{eq:B1}
\end{equation}
where
\begin{equation}
\underline{u}_i(t):=\max\{0,\ u_i(t)-\Delta i_{max}\},
\label{eq:B2}
\end{equation}
\begin{equation}
\bar{u}_i(t):=\min\left\{
u_i(t)+\Delta i_{max},\
\bar{u}_i^{v}(t),\
i_{ele,max,i}(T_{ele,i}(t)),\
\bar{u}_i^{p}(t)
\right\},
\label{eq:B3}
\end{equation}
where $\bar{u}_i^{v}(t)$ is the upper bound induced by the voltage constraint, and $\bar{u}_i^{p}(t)$ is the upper bound induced by the power constraint. Specifically,
\begin{equation}
\bar{u}_i^{v}(t):=
\frac{u_{ele,i}^{max}-u_{rev,i}}
{\rho_{1,i}+\rho_{2,i}T_{ele,i}(t)},
\label{eq:B4}
\end{equation}
\begin{equation}
\bar{u}_i^{p}(t)=
\frac{
-u_{rev,i}
+\sqrt{
u_{rev,i}^2
+
4(\rho_{1,i}+\rho_{2,i}T_{ele,i}(t))
P_{ele,max,i}(T_{ele,i}(t))
}
}{
2(\rho_{1,i}+\rho_{2,i}T_{ele,i}(t))
},
\label{eq:B5}
\end{equation}
where the nonnegative root is selected because the electrolyzer current is restricted to nonnegative values in the considered operating mode. 
Combining $\mathcal I_i(t)$ with the CBF-based HTO admissibility condition, the local admissible set is defined as
\begin{equation}\label{eq:B6}
\mathcal{U}_i(t):=
\left\{
u_i \in \mathcal{I}_i(t):
k_{1,i}(t)u_i^3-k_{2,i}(t)u_i^2+k_{3,i}(t)u_i+k_{4,i}(t)\ge 0
\right\}.
\end{equation}

If $\mathcal U_i(t)\neq\varnothing$, the minimum admissible current for electrolyzer $i$ is defined as
\begin{equation}
u_i^{-}(t):=\min\{u_i: u_i\in \mathcal U_i(t)\},
\label{eq:B7}
\end{equation}
and the minimum reachable total power is defined as
\begin{equation}
P_{min}^{reach}(t):=
\sum_{i=1}^{N_{ele}}P_{ele,i}(u_i^{-}(t),t).
\label{eq:B8}
\end{equation}

\textit{Proof.}
Assume that $\mathcal U_i(t)\neq\varnothing$ for all $i\in\{1,\dots,N_{ele}\}$.
Since $\mathcal I_i(t)$ is a compact interval and the polynomial inequality defining $\mathcal U_i(t)$ is continuous, $\mathcal U_i(t)$ is a closed subset of a compact set. Therefore, $\mathcal U_i(t)$ is compact and nonempty, and its minimum element $u_i^{-}(t)$ is well defined.
By definition, $u_i^{-}(t)\in\mathcal I_i(t)$, and therefore it satisfies the current limit, voltage limit, power limit, and ramp rate constraint. Moreover, since $u_i^{-}(t)\in\mathcal U_i(t)$, it also satisfies the CBF-based HTO constraint. Hence, the vector
\begin{equation}
u^{-}(t):=
\big[u_1^{-}(t),u_2^{-}(t),\dots,u_{N_{ele}}^{-}(t)\big]^\top,
\label{eq:B9}
\end{equation}
satisfies all individual constraints in (P2).
Since the electrolyzer power is monotonically increasing with respect to the current input over the admissible operating interval, $u_i^{-}(t)$ induces the minimum reachable power of electrolyzer $i$ under the admissibility constraints. Therefore, if
\begin{equation}
P_{min}^{reach}(t)\le P_{wind}(t),
\label{eq:B10}
\end{equation}
then $u^{-}(t)$ also satisfies the total power limit \eqref{E11b}. Therefore, $u^{-}(t)$ is a feasible point of (P2), which implies that the feasible set of (P2) is nonempty.
Finally, the objective function of (P2) is
\begin{equation}
J(u)=\frac12\|u-u_{des}(t+1)\|^2,
\label{eq:B11}
\end{equation}
which is continuous in $u$. Moreover, for each $i$, the constraint $u_i\in\mathcal I_i(t)$ restricts $u_i$ to a bounded interval, and thus the feasible set of (P2) is bounded. Since the CBF-based HTO constraints and the system level power limit are defined by continuous functions, the feasible set is also closed. Therefore, the feasible set of (P2) is compact. By the Weierstrass theorem, problem (P2) admits at least one optimal solution. \hfill$\blacksquare$

\bibliographystyle{elsarticle-num.bst}
\bibliography{reference.bib}
\end{document}